\documentclass[10pt,conference,english]{IEEEtran}
\usepackage[english]{babel}
\usepackage[utf8x]{inputenc}
\usepackage{times}

\usepackage[margin=18.5mm]{geometry}
\usepackage{xspace}
\usepackage{amsmath}
\usepackage{graphicx}
\usepackage[colorinlistoftodos]{todonotes}
\usepackage[normalem]{ulem}
\usepackage[colorlinks=true, allcolors=blue]{hyperref}
\usepackage{float}
\usepackage{amsbsy}
\usepackage{amssymb}
\usepackage{calrsfs}
\usepackage{multirow}

\newtheorem{theorem}{Theorem}
\newtheorem{lemma}{Lemma}

\newtheorem{proposition}{Proposition}
\newtheorem{example}{Example}
\newtheorem{corollary}[theorem]{Corollary}

\let\P\relax
\DeclareMathOperator\P{\textsf{P}}

\newcommand{\Xvh}{{\hat{\Xv}}}
\newcommand{\xv}{{\bf x}}
\newcommand{\xvh}{{\hat{\xv}}}
\newcommand{\Xv}{{\bf X}}

\newcommand{\yv}{{\bf y}}

\newcommand{\Yv}{{\bf Y}}

\newcommand{\wv}{{\bf w}}

\newcommand{\Mv}{{\bf M}}

\newcommand{\tv}{{\bf t}}

\newcommand{\oJ}{\overline{J}}

\newcommand{\Dv}{{\bf D}}

 \allowdisplaybreaks[4]

\title{On the Capacity Region for Secure Index Coding}
\author{
  Yuxin Liu$^*$, Badri N. Vellambi$^\dag$,  Young-Han Kim$^\ddag$, and Parastoo Sadeghi$^*$\\   
    $^*$Research School of Engineering, Australian National University, \{yuxin.liu, parastoo.sadeghi\}@anu.edu.au\\
  $^\dag$Department of Electrical Engineering and Computer Science, University of Cincinnati, badri.vellambi@uc.edu\\
  $^\ddag$Department of Electrical and Computer Engineering, University of California, San Diego, yhk@ucsd.edu
}

\newif\ifitw

\itwfalse

\begin{document}
\maketitle{}

\begin{abstract}
We study the index coding problem in the presence of an eavesdropper, where the aim is to communicate without allowing the eavesdropper to learn any single message aside from the messages it may already know as side information. We establish an outer bound on the underlying secure capacity region of the index coding problem, which includes polymatroidal and security constraints, as well as the set of additional decoding constraints for legitimate receivers. We then propose a secure variant of the composite coding scheme, which yields an inner bound on the secure capacity region of the index coding problem. For the achievability of secure composite coding, a secret key with vanishingly small rate may be needed to ensure that each legitimate receiver who wants the same message as the eavesdropper, knows at least two more messages than the eavesdropper. For all securely feasible index coding problems with four or fewer messages, our numerical results establish the secure index coding capacity region.
\end{abstract}

\section{Introduction}

Index coding is a canonical problem in network information theory with close connections to many important problems such as network coding \cite{effrosrouayheblangberg15} and distributed storage \cite{Shanmugam--Dimakis2014}. Index coding aims to find the optimal broadcast rate and optimal coding schemes for broadcasting $n$ unique messages from a server to $n$ receivers with (possibly differing) side information at each receiver \cite{birk1998informed}. Characterizing the capacity region of a general index coding problem remains elusive. This paper is concerned with a class of index coding problems where there is, in addition to $n$ legitimate receivers, an eavesdropper who may have side information about some messages and wants to obtain the rest. We aim to characterize inner and outer bounds on the secure index coding capacity region under the restricted security requirement that there is no leakage of information about any single message that is unknown to the eavesdropper.

The secure variant of the index coding problem was first studied in \cite{dau2012security}, where the conditions for a linear code to be a valid secure index code were investigated. Later in \cite{ong2016secure}, non-linear secure index codes that use side information as secret keys were proposed. The connection between secure network coding and secure index coding (analogous to the relationship between non-secure versions \cite{effrosrouayheblangberg15}) was developed in \cite{ong:kliewer:vellambi:isit2018}. In \cite{mojahedian2017perfectly}, the authors studied the minimum key length to achieve perfect secrecy where the eavesdropper has no additional side information, but it must not learn any information whatsoever about the messages (namely, zero mutual information). The private index coding problem with linear codes was studied in \cite{varun2018private} where the aim is to allow legitimate receivers to only learn about messages they want, but nothing of other unknown messages. Finally, \cite{fragouli:isit2017, fragouli:isit2018} considered the case in which the identity of the demanded message and the side information of each receiver should be kept private from other receivers. 

In this paper, we examine the fundamental limits of using side information as the main protection mechanism to effect security in the index coding problem. After introducing the system model and problem setup in Section \ref{sec:model}, Section \ref{sec:secure_outer} presents a newly developed outer bound on the secure index coding capacity region. Section \ref{sec:SCOMP} presents an achievable rate region using a secure random coding scheme for index coding. The proposed scheme is based on the existing composite coding scheme \cite{arbabjolfaei2013capacity,liu2017capacity}. For all securely feasible index coding problems with $n\leq 4$ messages, inner and outer bounds match, yielding  the corresponding secure capacity regions. However, we note that for the achievability of the secure composite coding scheme, a secret key with vanishingly small rate may be needed so that each legitimate receiver who wants the same message as the eavesdropper, knows at least two more messages than the eavesdropper.

\section{System Model} \label{sec:model}
\begin{figure} [b]
\begin{center}
\vspace{-3mm}
\includegraphics[scale=0.45]{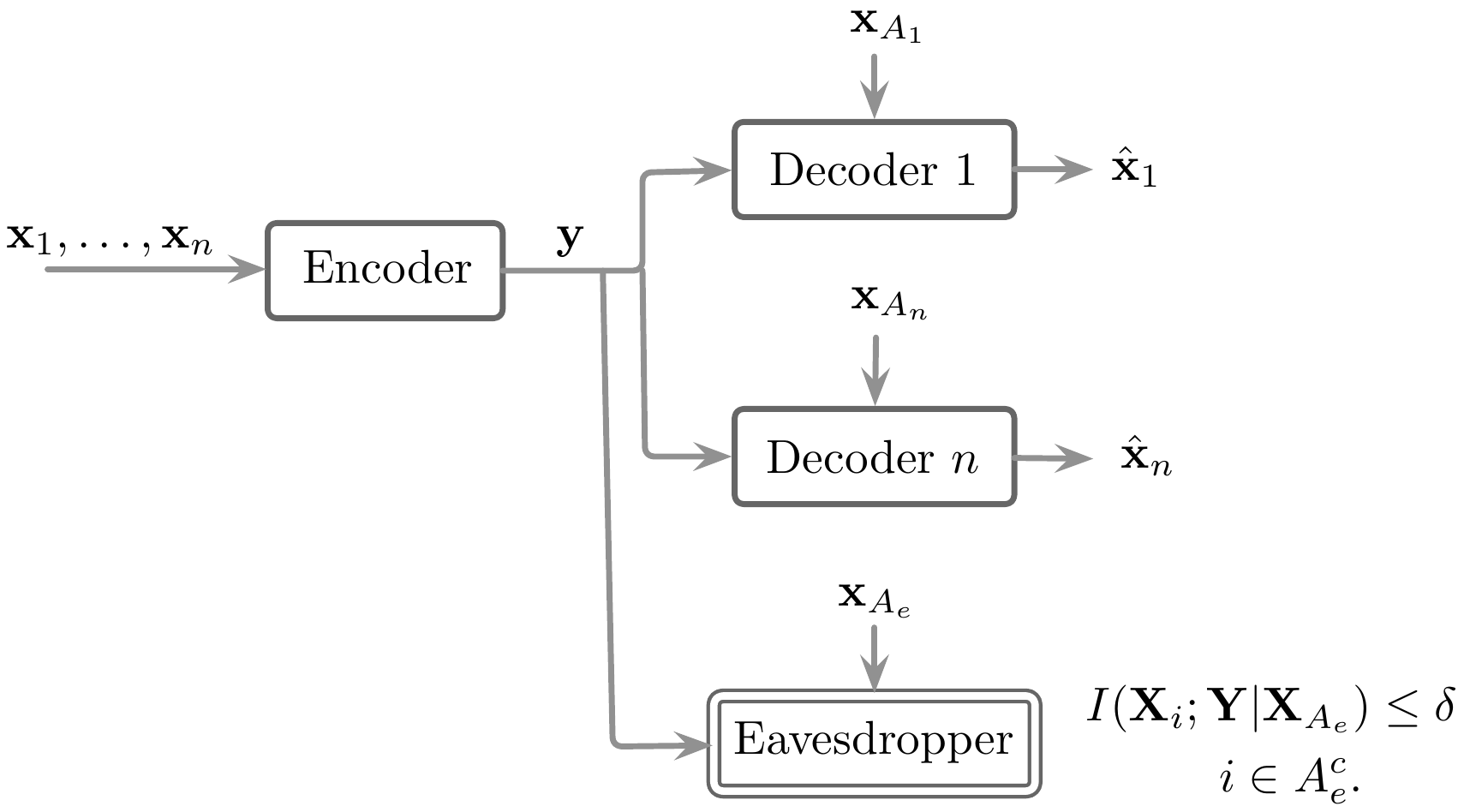}
\caption{Problem setup for secure index coding.}
\vspace{-3mm}
\label{fig:index_coding}
\end{center}
\end{figure}

Throughout the paper, we let $[n]\triangleq\{1,2,\dots,n\}$ and use $2^{[n]}$ to denote the power set of $[n]$. The aim of the index coding problem depicted in Figure \ref{fig:index_coding} is to devise a coding scheme to allow a server to communicate $n$ independent and uniformly distributed messages, $\Xv_i\in\{0,1\}^{t_i}$, $i\in [n]$, to their corresponding receivers over a noiseless broadcast link with unit capacity in the presence of an eavesdropper $e$. Each receiver has prior knowledge of the realization $\xv_{A_i}$ of $\Xv_{A_i}$, where $A_i \subseteq [n] \backslash \{i\}$. The set $B_i\triangleq [n]\setminus(A_i \cup \{i\})$ denotes the set of \emph{interfering} messages at receiver $i$. The eavesdropper has access to $\Xv_{A_e}, A_e \subset [n]$. The encoder should be designed to prevent the eavesdropper from learning any single message $\Xv_{j}$, $j\in A_e^c = [n] \backslash A_e$.

To compactly represent a non-secure index coding problem we use $(i|A_i), i\in [n]$, to indicate that legitimate receiver~$i$ has messages $\xv_{A_i}$ and wants to decode message $\xv_i$. With a slight abuse of notation, $(e|A_e)$ denotes what the eavesdropper has, but note that it wants to learn all other messages. A $(\tv,r) = ((t_1, \ldots, t_n), r)$ index code is defined by:
\begin{itemize}
\item One encoder at the server $\phi: \prod_{i=1}^n \{0,1\}^{t_i} \rightarrow \{0,1\}^r$ that uses all messages to generate the transmitted codeword $\Yv\triangleq\phi(\Xv_1,\ldots, \Xv_n)$ of length $r$ bits.
\item For each legitimate receiver $i\in [n]$, a decoder function $\psi_i: \{0,1\}^r \times \prod_{k\in A_i} \{0,1\}^{t_k} \rightarrow\{0,1\}^{t_i}$ that takes the received sequence $\Yv$ together with the side information at receiver $i$ and maps them to $\Xvh_i\triangleq\psi_i(\Yv,\Xv_{A_i})$.
\end{itemize}
A rate tuple $(R_1, \ldots, R_n)$ is said to be \emph{securely} achievable if for every $\delta,\epsilon >0$, there exists a $(\tv,r)$ index code such that the following three constraints are met:
\begin{align}
&\textrm{Rate: }\qquad\,\,\label{eq:rate_def} R_i \le \textstyle{\frac{t_i}{r}}, \quad i\in [n]; \\
&\textrm{Decoding: }\,\,\label{eq:error} \P\{(\Xvh_1,\ldots, \Xvh_n) \ne (\Xv_1, \ldots, \Xv_n)\} \le \epsilon;\\
&\textrm{Security: }\,\,\,\,\,\,\label{eq:secu} I(\Xv_i; \Yv |\Xv_{A_e}) \le \delta,  \quad i \in A_e^c.
\end{align}
We define the secure capacity region $\mathcal{C}^{\mathrm{s}}$ as the closure of the set of all securely achievable rate tuples. Note that for a sequence of codes operating at a securely achievable rate tuple, the decoding condition along with Fano's inequality ensure that at receiver $i$, $\lim_{\epsilon\to0} H(\Xv_i|\Yv,\Xv_{A_i})= 0$.

Note that an index coding problem is not securely feasible when there is there is no securely achievable rate tuple. This happens when $A_i \subseteq A_e$ for some  $i \in A^c_e$. That is, when the side information of the eavesdropper is equally strong or stronger than that of some receiver.  Otherwise, the secure index coding problem is said to be \emph{securely feasible}.

\section{Polymatroidal Outer Bound} \label{sec:secure_outer}
We present the following outer bound to the secure index coding  capacity region.

\begin{theorem} [Secure Outer Bound]\label{theo:SPM}
Any securely achievable rate tuple for the index coding problem $(i|A_i)$, $i \in [n]$, and $(e|A_e)$ must lie in $\mathcal{R}_{\mathrm{g}}^\mathrm{s}$ that consists of all rate tuples satisfying
\begin{align}\label{eq:R:sec}
R_i = g(B_i \cup \{i\})-g(B_i), ~~i \in [n],
\end{align}
for some set function $g: 2^{[n]}\rightarrow [0,1]$ such that for any $J\subseteq [n]$ and $i,k\notin J$,
\begin{align}
\label{pm_as_1} &g(\emptyset) = 0,\\
\label{pm_as_2} &g([n]) \leq 1, \\
\label{pm_as_3} &g(J) \leq g(J,\{i\}), \\
\label{pm_as_4} &g(J) + g(J\cup \{i,k\}) \leq g(J \cup \{i\}) + g(J \cup \{k\}),\\
\label{pm_as_5} &g(B_i \cup \{i\}) - g(B_i) = g(\{i\}),
\end{align}
and additionally for $i\in A_e^c$,
\begin{align}
\label{item:security}  &g(A_e^c\setminus \{i\}) \geq g(A_e^c).
\end{align}
$\hfill \blacksquare$
\end{theorem}
The proof is given in Appendix \ref{proof:SPM}. 
A few remarks are in order. First, we note that the rate constraint \eqref{eq:R:sec} and polymatroidal constraints \eqref{pm_as_1}-\eqref{pm_as_4} appeared in \cite{arbabjolfaei2013capacity} to form an outer bound on the non-secure index coding capacity region\footnote{In \cite{arbabjolfaei2013capacity}, inequalities in \eqref{eq:R:sec} and equality in \eqref{pm_as_2} were used. This is immaterial to the non-secure capacity region outer bound. See Remark \ref{rem:eqaulaity} at the end of this section for its impact on the secure counterpart.}, which is known to be tight for all index coding problems with $n \le 5$ messages. Constraint \eqref{pm_as_5} captures \emph{additional decoding conditions} for each legitimate receiver $i \in [n]$, since
\begin{align}
H(\Xv_i|\Yv,\Xv_{A_i}) = H(\Xv_i|\Yv,\Xv_{A_i},\Xv_{C}) = 0,
\end{align}
for any $C \subseteq B_i$. Due to the submodularity constraint \eqref{pm_as_4}, it suffices to write the additional decoding condition for $C = B_i$ only. See Appendix \ref{proof:SPM} for more details. We note that the same constraint appeared in a similar outer bound to the non-secure index coding capacity region in \cite{lexico}. Finally, \eqref{item:security} captures the security constraint \eqref{eq:secu} with $\delta = 0$.

An \textit{explicit} outer bound to the index coding capacity region is derived from Theorem \ref{theo:SPM} by the means of Fourier-Motzkin elimination (FME)~\cite{elgamal_yhk} through eliminating $g(J), J\subseteq [n]$ that are viewed purely as intermediate variables.  Let us consider a non-secure and a secure example.

\begin{example} \label{exp:NonSPM}
Consider the \emph{non-secure} index coding problem $(1|-), (2|3), (3|2)$ in the absence of the eavesdropper. Invoking Theorem \ref{theo:SPM} without \eqref{item:security} and eliminating variables $g(J), J\subseteq [n]$, via FME yields
\begin{align}
R_1 + R_2 \leq 1, ~~~R_1 + R_3 \leq 1,
\end{align}
which is the explicit outer bound on the non-secure index coding capacity region. $\hfill \blacksquare$
\end{example}

\begin{example}  \label{exp:SPM}
Consider the index coding problem $(1|-), (2|3), (3|2)$, with the eavesdropper~$(e|1)$. The explicit outer bound on the secure capacity region derived from Theorem \ref{theo:SPM} is
\begin{align*}
\qquad\qquad\qquad R_2 = R_3, ~~ R_1+R_3 \leq 1. \qquad\qquad\qquad\quad \blacksquare
\end{align*}
\end{example}

In Example \ref{exp:SPM}, the security requirement imposes the equality $R_2 = R_3$. Since the eavesdropper already has $\xv_1$ as side information, it is not possible to protect $\xv_2$ or $\xv_3$ using $\xv_1$. Therefore, the only solution to guarantee secrecy is to protect $\xv_2$ with $\xv_3$ and vice versa at the same rate. This can be achieved using a simple linear code, $\xv_1$ (of length $t_1$), $\xv_2 \oplus \xv_3$ (of length $t_2$), illustrated below.
\begin{figure} [b]
\vspace{-8mm}
\begin{center}
\includegraphics[scale=0.9]{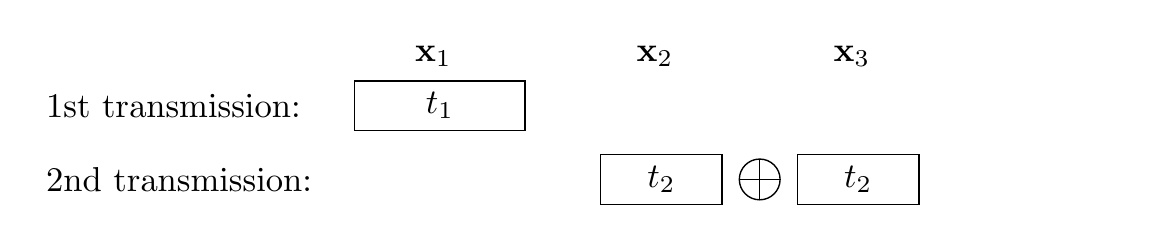}
\vspace{-1mm}
\caption{A linear code achieves the capacity region for Example \ref{exp:SPM}.}
\vspace{-1mm}
\label{fig:exp2_code}
\end{center}
\end{figure}

We summarize a few important observations.
\begin{enumerate}
\item The outer bound on the non-secure capacity region of the index coding problem is at least as large as that on the secure capacity region. 
\item In some secure index coding problems, it is possible for a stronger receiver (with more side information) to have its rate bounded by that of a weaker receiver (with less side information). Equivalently, it is possible that
\begin{equation}
\nonumber  \exists i,j \in[n] \text{ such that } A_i \subset A_j, R_i \geq R_j.
\end{equation}
For the problem $(1|3),(2|1,3),(3|1),(e|-)$, the outer bound on the secure capacity region stipulates $R_2 \leq R_3$, while $A_3 =\{3\} \subset \{1, 3\} = A_2$.
\item The additional decoding constraint \eqref{pm_as_5} in Theorem \ref{theo:SPM} is essential for deriving a tighter secure capacity region outer bound for some index coding problems. For example, if we exclude \eqref{pm_as_5}, the outer bound for the index coding problem $ (1|3),(2|3),(3|2), (e|-)$ is
\begin{equation}
\nonumber R_1+R_2\leq 1, \quad R_2 = R_3.
\end{equation}
However, with  \eqref{pm_as_5} included, the outer bound is
\begin{equation}
\nonumber R_1+R_2\leq 1, \quad R_2 = R_3, \quad R_1 \leq R_3.
\end{equation}
\item To obtain equality relationships between the message rates, $R_i = g(B_i \cup \{i\})-g(B_i)$ should be used in Theorem \ref{theo:SPM}, instead of $R_i \leq g(B_i \cup \{i\})-g(B_i)$. For example, if the latter is used, then $R_2 = R_3$  will not be captured in Example \ref{exp:SPM}. To ensure the convex envelope of the capacity outer bound is obtained, we then use $g([n]) \leq 1$ in Theorem \ref{theo:SPM}, instead of $g([n]) = 1$.\label{rem:eqaulaity}
\end{enumerate}

\section{Secure Composite Coding Inner Bound} \label{sec:SCOMP}

Before proposing the secure composite coding scheme, we first recap the original composite coding scheme established in \cite{arbabjolfaei2013capacity}. For ease of exposition, the scheme is described for a fixed decoding configuration, which is a tuple of subsets of messages $\Dv = (D_i, i\in[n])$ such that for each $i\in [n]$, $D_i\subseteq [n] \backslash A_i$ and $i\in D_i$. Let $r \in \mathbb{N}$. Let for each $i \in [n]$, $t_i = \lceil{rR_i\rceil}$. Denote $s_K = \lceil{rS_K\rceil}$, where $S_K\in[0,1]$ is the rate of composite index for subset $K$. By convention, $S_\emptyset = 0$.

\textbf{Codebook generation:} \textbf{(1)} For each $K \subseteq [n]$ and $\xv_K$, a corresponding composite index $W_{K}(\xv_K)$ is drawn uniformly at random from $[2^{s_K}]$. \textbf{(2)} For every tuple $(w_{K}, K\in 2^{[n]})$, the codeword to be transmitted, $\Yv((w_{K}, K\in 2^{[n]}))$, is drawn uniformly at random from $[2^r]$.  The random codebooks (message-to-composite indices and composite indices-to-codeword maps) are revealed to all parties. 

\textbf{Encoding:} To communicate a realization  $\xv_{[n]}$, the transmitter sends $\Yv((W_{K}(\xv_K), K\in 2^{[n]}))$.

\textbf{Decoding:} Upon receiving the codeword realization $\mathbf y$:
 \textbf{(1)}  Each legitimate receiver $i$ finds the unique tuple of composite indices $(\hat{w}_{K}, K\in 2^{[n]})$ such that $\yv = \Yv((\hat{w}_{K}, K\in 2^{[n]}))$, and declares an error if a unique tuple is not found. \textbf{(2)} Assuming composite index tuple decoding is successful, receiver $i$ finds the unique message tuple $\xvh$ such that $w_{K} = W_{K}(\xvh_K)$, for all $K \subseteq D_i \cup A_i$. An error is declared if a unique tuple is not found.

The following result from \cite{arbabjolfaei2013capacity} quantifies the constraints on the message rates and composite index rates for successful decoding (in the non-secure setting).

\begin{proposition}\label{thm:compcod}
A rate tuple $(R_i, i\in [n])$
is achievable for the index coding problem $(i| A_i)$, $i \in [n]$, if for each $i\in [n]$:
\begin{align}
\label{deco_1}  &\sum_{J \not\subseteq A_i} S_J < 1,\\
\label{deco_2} &\sum_{i \in K} R_i < \sum_{J \subseteq D_i \cup A_i : J \cap K \neq \emptyset} S_J, \quad K \subseteq D_i. 
\end{align}
\end{proposition}

Now we move on to develop the \emph{secure} composite coding scheme.
Recall the security condition
\begin{align}
\label{secure_eq} I(\Xv_i ; \Yv|\Xv_{A_e}) < \delta, ~~ i \in A_e^c.
\end{align}
Using the chain rule for mutual information  and the independence between different messages we have
\begin{align}\label{secure_eq_new}
I(\Xv_i; \Yv, \Xv_{A_e})  < \delta, ~~ i \in A_e^c.
\end{align}
Since the eavesdropper can generate all composite indices $\{\wv_J: J \subseteq  A_e\}$ from $\Xv_{A_e}$, it will be useful to define $T = \{K: K \subseteq [n], K \not\subseteq A_e\}$. Then for any $Q \subseteq T$, $P_Q = \bigcup_{J \in Q} J \backslash A_e$ is the set of messages from $Q$ that are unknown to the eavesdropper. We assume that the eavesdropper learns the codebook and is also able to decode all the composite indices in the first step of decoding. Condition \eqref{secure_eq_new} becomes: 
\begin{align}
\label{final_secure_eq2} I(\Xv_i ; \{W_K: K \in T\}, \Xv_{A_e}) &< \delta, ~~ i \in A_e^c.
\end{align}

Applying Theorem 1 from \cite{6283010} and Lemma 2.7 from \cite{csiszar2011information}, we obtain the following random-coding based achievable rate region. The proof will be provided in Appendix \ref{proof:SCOMP} for the more general secure enhanced composite coding scheme described in Proposition \ref{theo:SCOMPenhanced}.
\begin{theorem} \label{theo:SCOMP}
A rate tuple $(R_i, i\in [n])$
is securely achievable for the index coding problem $(i|A_i), i\in [n], (e|A_e)$ if
\begin{align}
\label{deco_1s} & \sum_{J \not\subseteq A_i} S_J< 1, \quad i\in [n],\\
\label{deco_2s} &\sum_{i \in K} R_i< \sum_{\substack{J \subseteq D_i \cup A_i \\ J \cap K \neq \emptyset}} S_J,  \quad K \subseteq D_i, i\in [n],\\
\label{eq:sec_comp} &\sum_{\substack{K \subseteq P_Q\cup A_e\\ K \not\subseteq A_e}} S_K <  \sum_{j \in (P_Q\backslash \{i\})} R_j, \quad Q \subseteq T, i \in A_e^c.
\end{align}
\end{theorem}

Note, when Theorem \ref{theo:SCOMP} gives an inequality of the form $S_J <0$, we set $S_J = 0$. For each index coding problem with  $n\leq5$ messages, a single \emph{natural} decoding configuration $\underline \Dv$ \cite{liusimplified} was shown to be sufficient to achieve the non-secure  capacity region. We  will also use the natural decoding configuration in this paper, which will be sufficient to achieve the secure capacity region for all index coding problems with $n \leq 4$ messages. However, more than one decoding configuration might be necessary for larger problems. Secure composite coding with multiple decoding configurations is detailed below. 

\subsubsection{Secure Enhanced Composite Coding Scheme}  Following similar lines as \cite{liu2017capacity}, let $\Delta$ be the set of all decoding configurations, i.e, $\Delta = \{\Dv: D_i\subseteq [n] \backslash A_i,  i\in D_i\}$.

Let $r \in \mathbb{N}$. Let for each $\Dv\in\Delta$ and $i \in [n]$, $t_i(\Dv) = \lceil{rR_i(\Dv)\rceil}$, where $R_i(\Dv)$ is the rate of message $i$ communicated via decoding configuration $\Dv$. Let $\Xv_i(\Dv)\in [2^{t_i(\Dv)}]$ be the part of message $i$ communicated via decoding configuration $\Dv$. For each $K \subseteq [n]$ and $\Dv\in \Delta$, let $S_K(\Dv)\in[0,1]$. Denote $s_K(\Dv) = \lceil{rS_K(\Dv)\rceil}$, $K \subseteq [n]$, where $S_K(\Dv)$ is the rate of composite index for subset $K$ and configuration $\Dv$. By convention, $S_\emptyset(\Dv) = 0$ for each $\Dv\in\Delta$.

\textbf{Codebook generation:} \textbf{(1)} For each $K \subseteq [n]$, $\Dv\in\Delta$, and $\xv_K(\Dv)$, a corresponding composite index $W_{K,\Dv}(\xv_K(\Dv))$ is drawn uniformly at random from $[2^{s_K(\Dv)}]$. \textbf{(2)} For every tuple $(w_{K,\Dv}, (K,\Dv)\in 2^{[n]}\times \Delta)$, the codeword to be transmitted, $\Yv((w_{K,\Dv}, (K,\Dv)\in 2^{[n]}\times \Delta))$, is drawn uniformly at random from $[2^r]$.  The random codebooks (message-to-composite indices and composite indices-to-codeword maps) are revealed to all parties. 

\textbf{Encoding:} To communicate a realization  $\xv_{[n]}$, the transmitter sends $\Yv((W_{K,\Dv}(\xv_K(\Dv)), (K,\Dv)\in 2^{[n]}\times \Delta))$.

\textbf{Decoding:} Upon receiving the codeword realization $\mathbf y$:
 \textbf{(1)}  Each legitimate receiver $i$ finds the unique tuple of composite indices $(\hat{w}_{K,\Dv}, (K,\Dv)\in 2^{[n]}\times \Delta)$ such that $\yv = \Yv((\hat{w}_{K,\Dv}, (K,\Dv)\in 2^{[n]}\times \Delta))$, and declares an error if a unique tuple is not found.  \textbf{(2)} Assuming composite index tuple decoding is successful, for each $\Dv\in\Delta$, receiver $i$ finds the unique message tuple $\xvh_{D_i}(\Dv)$ such that $w_{K,\Dv} = W_{K,\Dv}(\xvh_K)$, for all $K \subseteq D_i \cup A_i$. An error is declared if a unique tuple is not found.
\begin{proposition} \label{theo:SCOMPenhanced}
A rate tuple $(R_i, i\in [n])$
is securely achievable for the index coding problem $(i|A_i), i\in [n], (e|A_e)$ if
\begin{align}
&R_i = \sum_{\Dv \in \Delta} R_i(\Dv), \quad i\in [n], \\
\label{deco_1s} &\sum_{\Dv \in \Delta} \sum_{J \not\subseteq A_i} S_J(\Dv) < 1, \quad i\in [n],\\
\label{deco_2s} &\sum_{i \in K} R_i(\Dv) < \sum_{\substack{J \subseteq D_i \cup A_i \\ J \cap K \neq \emptyset}} S_J(\Dv),  \quad K \subseteq D_i, i\in [n],\\
\label{eq:sec_comp} &\sum_{\substack{K \subseteq P_Q\cup A_e\\ K \not\subseteq A_e}} S_K(\Dv) <  \sum_{j \in (P_Q\backslash \{i\})} R_j(\Dv), \quad Q \subseteq T, i \in A_e^c.
\end{align}
\end{proposition}

Note that in Proposition \ref{theo:SCOMPenhanced}, if we set $S_K(\Dv)$, $K \in 2^{[n]}$ and $R_j(\Dv)$, $j \in [n]$ to zero for all, but one particular $\Dv$, we will recover Theorem \ref{theo:SCOMP}.

\subsection{Secure Composite Coding with a Secret Key}

Theorem \ref{theo:SCOMP} and Proposition \ref{theo:SCOMPenhanced} may generate conflicting constraints for some index coding problems as shown below. 
\begin{example}  \label{exp:SCOMP}
Consider the same setting as in Example \ref{exp:NonSPM}. Set $D_1 = \{1\}$, $D_2 = \{1,2\}$, and $D_3 = \{1,3\}$. The set of active inequalities generated by Theorem \ref{theo:SCOMP} is
\begin{align*}
    &S_1+S_2+S_{12}+S_3+S_{13}+S_{23}+S_{123} < 1, \notag \\
        &R_1 < S_1,\notag\\    
    &R_2 < S_2+S_{12}+S_{23}+S_{123}, \notag\\
    &R_3 < S_3+S_{13}+S_{23}+S_{123},\\
    &R_1+R_2 < S_1+S_2+S_{12}+S_{13}+S_{23}+S_{123},\notag \\
    &R_1+R_3 < S_1+S_{12}+S_3+S_{13}+S_{23}+S_{123},\notag \\
    &S_2+S_{12}+S_{13}+S_{23}+S_{123} < R_2,\\
    &S_{12}+S_3+S_{13}+S_{23}+S_{123} < R_3,\\
    ~&S_2 =0, ~  S_3 = 0,  ~S_{12}=0,~S_{13}=0.  \notag 
\end{align*}
Clearly, there are conflicting constraints for $R_2$ and $R_3$. $\hfill\blacksquare$
\end{example}

For $n=3$ messages, there are the total of 20 securely feasible index coding problems. Of these, only the problem $(1|2,3), (2|1,3), (3|1,2), (e|-)$ does not have conflicting inequalities. For $n=4$ messages, 43 out of 833 securely feasible index coding problems have no conflicting inequalities. For all such non-conflicting cases, the secure composite coding inner bound matches the secure polymatroidal outer bound, thereby establishing the corresponding secure capacity region. In each of these problems, each receiver who wants the same message as the eavesdropper, knows at least two more messages as side information than the eavesdropper. I.e., $\forall \, i \in A_e^c, |A_i \backslash A_e| \geq 2$.

We now show a resolution can be obtained for the conflicting cases by using a secret key of arbitrarily small rate shared  between the server and legitimate receivers.

Assume there is an independent secret key $\Mv$ at rate $\zeta_M$, shared between the transmitter and all the legitimate receivers. For each $K \subseteq [n]$, $\xv_{K}(\Dv) \cup \Mv$ is mapped into a composite index $W_{K,M, \Dv,}(\xv_K(\Dv)\cup \Mv)$ drawn uniformly at random from $[2^{s_{K,M}(\Dv)}]$. The second step of codebook generation, encoding and decoding are the same as before.
\begin{theorem} \label{theo:SCOMP_KEY}
A rate tuple $(R_i, i\in [n])$
is securely achievable for the index coding problem $(i|A_i), i\in [n], (e|A_e)$ if
\begin{align}
&R_i = \sum_{\Dv \in \Delta} R_i(\Dv), \quad i\in [n], \\
\label{deco_1skey} &\sum_{\Dv \in \Delta} \sum_{J \not\subseteq A_i} S_{J,M}(\Dv) < 1, \quad i\in [n],\\
\label{deco_2skey} &\sum_{i \in K} R_i(\Dv) < \sum_{\substack{J \subseteq D_i \cup A_i \\ J \cap K \neq \emptyset}} S_{J,M}(\Dv),  K \subseteq D_i, i\in [n],\\
\label{eq:sec_compskey} &\sum_{\substack{K \subseteq P_Q \cup A_e\\ K \not\subseteq A_e}} S_{K,M}(\Dv) <  \sum_{j \in (P_Q\backslash \{i\})} R_j(\Dv)+\zeta_M,\\\nonumber &\quad\quad \quad Q \subseteq T, i \in A_e^c.
\end{align}

\end{theorem}

For the secure index coding problem described in Example \ref{exp:SCOMP}, the secure achievable rate region with a secret key becomes
\begin{align}
&R_3 - \zeta_M < R_2 < R_3 + \zeta_M, \notag \\
&R_1+ R_2 < 1 , \quad R_1 + R_3  <1. \notag
\end{align}
which matches the polymatroidal outer bound as $\zeta_M \rightarrow 0$.

We now summarize our key observations.
\begin{enumerate}
\item For cases where there are at least two more messages at each receiver $i$, $i\in A_e^c$ to protect their desired message (i.e., $\forall\, i\in A_e^c, |A_i \backslash A_e| \geq 2$), the proposed secure composite coding achieves capacity without the need for a shared secret key. For $n=3$, there is 1 such problem out of 20 securely feasible problems. For $n=4$, there are 43 such problems out of 833 securely feasible problems. 
\item For remaining cases, conflicting inequalities can be resolved by means of a secret key of vanishingly small rate. The secret key acts as the second message unknown to the eavesdropper to ensure $\forall\, i\in A_e^c, |A_i \backslash A_e| \geq 2$. 
\ifitw
\item Table \ref{table:SPM-3} lists the secure capacity region for 20 securely feasible index coding problems with $n=3$ messages.
\else
\item Appendix \ref{app:table3} lists the secure capacity region for all 20 securely feasible index coding problems with $n=3$ messages.
\fi
\end{enumerate}

\ifitw
\else

\section{Equivalence between Two Capacity Region Outer Bounds} \label{sec:equi}
First, let us specialize  Theorem \ref{theo:SPM} to the non-secure index coding.
\begin{corollary} [Non-secure Outer Bound]\label{theo:PM}
Any achievable rate tuple for the index coding problem $(i|A_i)$, $i \in [n]$ must lie in $\mathcal{R}_{\mathrm{g}}$ that consists of all rate tuples satisfying
\begin{align}\label{pm_a_0}
R_i \leq g(B_i \cup \{i\})-g(B_i), ~~i \in [n],
\end{align}
for some set function $g: 2^{[n]}\rightarrow [0,1]$ such that for any $J\subseteq [n]$ and $i,k\notin J$,
\begin{align}
\label{pm_a_1} &g(\emptyset) = 0,\\
\label{pm_a_2} &g([n]) = 1, \\
\label{pm_a_3} &g(J) \leq g(J,\{i\}), \\
\label{pm_a_4} &g(J) + g(J\cup \{i,k\}) \leq g(J \cup \{i\}) + g(J \cup \{k\}),\\
\label{pm_a_5} &g(B_i \cup \{i\}) - g(B_i) = g(\{i\}).
\end{align}
$\hfill \blacksquare$
\end{corollary}
We have used inequalities in  \eqref{pm_a_0} and equality in \eqref{pm_a_2}.\footnote{This is  technically needed in the proof of Theorem \ref{Prop_hg_equi}, but it is immaterial to the outer bound itself.} 

Let  $\Xv_0 =\Yv$ denote a random variable over $\{0,1\}^r$ representing the output of the index code. Denote $N = \{0\} \cup[n]$ and define the entropic set function $h: 2^{\{0\}\cup [n]} \rightarrow R_{\geq 0}$ as
\begin{align}
  h(J) = H(\Xv_J).
  \end{align}
The following is the outer bound on the non-secure capacity region of the index coding that captures all Shannon-type inequalities of the entropy function.
\begin{theorem} \label{thm:h:outerbound}Any achievable rate tuple for the index coding problem $(i|A_i)$, $i \in [n]$ must lie in $\mathcal{R}_{\mathrm{h}}$ consisting of all rate tuples $(R_i, i\in [n])$ that satisfy
\begin{align}
\label{SH_rate}  R_i \leq \frac{h(\{i\})}{h(\{0\})}, ~~i \in [n],
\end{align}
for  some set function $h: 2^{\{0\}\cup [n]} \rightarrow R_{\geq 0}$ such that
\begin{align}
\label{SH_1} &h(\emptyset) = 0, \\
\label{SH_2} &h([n]) = \sum_{i\in [n]} h(\{i\}), \\
\label{SH_4} &h(N \setminus \{i\}) = h(N) = 1,~~i\in [n],\\
\label{SH_5} &h(\{i\} \cup A_i \cup \{0\}) = h(A_i \cup \{0\}), ~~i\in [n], \\
\label{SH_6} &h(J) \leq h(J\cup \{i\}),\quad J\subseteq N, i\in N\backslash J,\\
\label{SH_7} &h(J) + h(J \cup \{i,k\}) \leq h(J \cup \{i\}) + h(J \cup \{k\}),\\
&\phantom{{} = ~~~~} J\subseteq N,i,k \not\in J, i\neq k. \notag
\end{align}
\end{theorem}

We now state the main result of this section.  

\begin{theorem} \label{Prop_hg_equi}
$\mathcal{R}_{\mathrm{h}}  = \mathcal{R}_{\mathrm{g}}$.
\end{theorem}

The proof of Theorem \ref{Prop_hg_equi} is shown in detail in Appendix \ref{proof:g_to_h}, and the outline is presented here.
\begin{itemize}
\item To prove $\mathcal{R}_{\mathrm{g}} \subseteq \mathcal{R}_{\mathrm{h}}$, we first take a set function $g$ which satisfies \eqref{pm_a_0} to \eqref{pm_a_5}. We then define
 $h: 2^{\{0\} \cup [n]} \rightarrow R_{\geq 0}$ as follows. For $J \subseteq [n], i\in [n]$,
\begin{align}
\label{g_to_h_1} h(\emptyset) &= 0, \\
\label{g_to_h_2} h(\{i\}) &= \frac{g(\{i\})}{\sum_{i=1}^n g(\{i\})}, \\
\label{g_to_h_3} h(J) &= \sum_{i \in J} h(\{i\}),\\
\label{g_to_h_4} h(J \cup \{0\}) &= h(J)+\frac{g(\overline{J})}{\sum_{i=1}^n g(\{i\})},
\end{align}
where for notational convenience, we use $\oJ$ to denote $J^c = [n]\setminus J$. We then prove \eqref{SH_rate}-\eqref{SH_7} using the constraints of Corollary \ref{theo:PM}.
\item To prove  $\mathcal{R}_{\mathrm{h}} \subseteq \mathcal{R}_{\mathrm{g}}$, we take a set function $h$ that satisfies \eqref{SH_rate} to \eqref{SH_7} and define
\begin{align}\label{h_to_g}
g(J) = \frac{h(\overline{J} \cup \{0\}) - h(\overline{J})}{h(\{0\})}, \quad J \subseteq [n].
\end{align}
We then prove \eqref{pm_a_0} to \eqref{pm_a_5} using the constraints of Theorem \ref{thm:h:outerbound}.
\end{itemize}

\fi

\newcommand{\noopsort}[1]{}

\ifitw

\else

\appendices

\section{Proof of Theorem \ref{theo:SPM}} \label{proof:SPM}

For the sake of simplicity, we assume perfect decoding, i.e., $H(\Xv_i|\Yv,\Xv_{A_i})=0$. This allows us to ignore infinitesimal terms that will appear if we use the actual decoding constraint in conjunction with Fano's inequality. We also assume $\delta = 0$.

We start with \eqref{eq:rate_def} and use message independence and  the decoding constraint  as follows.
\begin{align}
R_i &\leq \frac{t_i}{r}  = \frac{1}{r} H(\Xv_i) \\
&= \frac{1}{r}\big(H(\Xv_i|\Xv_{A_i}) - H(\Xv_i|\Yv, \Xv_{A_i})\big) \\
&= \frac{1}{r}I(\Xv_i ; \Yv|\Xv_{A_i}) \\
&= \frac{1}{r} H(\Yv|\Xv_{A_i}) - \frac{1}{r} H(\Yv|\Xv_i, \Xv_{A_i}) \\
&= g(B_i \cup \{i\}) - g(B_i),\label{eq:g_bi}
\end{align}
where the set function $g$ is defined as 
\begin{align}
g(J) = \frac{1}{r} H(\Yv|\Xv_{J^c}),  \quad J \subseteq [n].
\end{align}
The first four constraints follow from the definition of the set function $g$ and basic properties of  the entropy function.

\begin{itemize}
\item To prove \eqref{pm_as_5}, recall that
\begin{align}
\label{addi_eq} H(\Xv_i | \Yv, \Xv_{A_i}) = H(\Xv_i | \Yv,\Xv_{A_i}, \Xv_{C_i})  = 0,
\end{align}
for all $C_i \subseteq B_i$, as conditioning cannot increase entropy. Now let $C_i = B_i$, $i \in [n]$ and write
\begin{align}
\nonumber H(\Xv_i) &=H(\Xv_i|\Xv_{A_i}, \Xv_{B_i}) -H(\Xv_i | \Yv,\Xv_{A_i}, \Xv_{B_i})\\\nonumber&= I(\Xv_i;\Yv|\Xv_{A_i}, \Xv_{B_i}) \\\nonumber
&=H(\Yv|\Xv_{A_i}, \Xv_{B_i}) -H(\Yv | \Xv_i,\Xv_{A_i}, \Xv_{B_i}) \nonumber\\&= H(\Yv|\Xv_{A_i}, \Xv_{B_i}) = rg(\{i\}),\label{eq:g_i}
\end{align}
where the last equality follows from $H(\Yv|\Xv_{[n]}) = 0$ and the definition of $g$. Comparing \eqref{eq:g_bi} and \eqref{eq:g_i} proves
\begin{align*}
g(\{i\}) = g(B_i\cup\{i\}) - g(B_i).
\end{align*}
Due to the submodularity of $g$
\begin{align*}
g(\{i\})\geq g(C_i\cup\{i\}) - g(C_i)  \geq  g(B_i\cup \{i\})- g(B_i),
\end{align*}
for all $C_i \subset B_i$. Hence, adding $g(\{i\}) = g(B_i\cup\{i\}) - g(B_i)$ to the constraints will suffice.

\item To prove \eqref{item:security}, rewrite  \eqref{eq:secu}  as
\begin{align}
I(\Xv_i; \Yv| \Xv_{A_e}) &= 0\\
\Rightarrow\quad H(\Yv|\Xv_{A_e}) &= H(\Yv|\Xv_i, \Xv_{A_e}) \\
\Rightarrow\quad g(A_e^c) &= g([n] \backslash (\{i\} \cup A_e)).
\end{align}
Hence, noting the monotonicity condition \eqref{pm_as_3}, when $J = [n]\backslash (\{i\} \cup A_e)$, \eqref{item:security} holds. 
\end{itemize}

\section{Secure Capacity Region of the Index Coding Problems with $n=3$ Messages}\label{app:table3}
Table \ref{table:SPM-3} summarizes the results. Aside from the problems shown in Table \ref{table:SPM-3}, all other problems are securely infeasible.
\fi
\begin{table}[tbh]
\caption {Secure capacity region outer bounds for all securely feasible index coding problems with $n=3$ messages} \label{table:SPM-3} 

\begin{tabular}{|c|c|c|c|c|c|c|c|c|}
\hline
\textbf{\begin{tabular}[c]{@{}c@{}}Receiver and \\ Eavesdropper\\ Information\end{tabular}} & \textbf{\begin{tabular}[c]{@{}c@{}}Outer \\ Bounds\end{tabular}}                                                             & \textbf{\begin{tabular}[c]{@{}c@{}}Receiver and \\ Eavesdropper\\ Information\end{tabular}} & \textbf{\begin{tabular}[c]{@{}c@{}}Outer \\ Bounds\end{tabular}}                            \\ \hline
\begin{tabular}[c]{@{}c@{}}$(1|-)$\\ $(2|3)$\\ $(3|2)$\\ $(e|1)$\end{tabular}                       & \begin{tabular}[c]{@{}c@{}}$R_2=R_3$\\$ R_1+R_3 \leq 1$\end{tabular}                                                                  & \begin{tabular}[c]{@{}c@{}}$(1|2,3)$\\$ (2|1)$\\$ (3|-)$\\ $(e|3)$\end{tabular}                     & \begin{tabular}[c]{@{}c@{}}$R_2 + R_3 \leq 1$\\ $R_1 = R_2$\end{tabular}                                \\ \hline
\begin{tabular}[c]{@{}c@{}}$(1|3)$\\$ (2|3)$\\ $(3|2)$\\$ (e|-)$\end{tabular}                       & \begin{tabular}[c]{@{}c@{}}$R_1 + R_2 \leq 1$\\ $R_2 =R_3$\\$ R_1 \leq R_3$\end{tabular}                                                   & \begin{tabular}[c]{@{}c@{}}$(1|3)$\\ $(2|3)$\\ $(3|2)$\\ $(e|1)$\end{tabular}                       & \begin{tabular}[c]{@{}c@{}}$R_1 + R_2 \leq 1$\\$ R_2 =R_3$\end{tabular}                                 \\ \hline
\begin{tabular}[c]{@{}c@{}}$(1|3)$\\$ (2|1)$\\ $(3|2)$\\ $(e|-)$\end{tabular}                       & \begin{tabular}[c]{@{}c@{}}$R_1 + R_2 \leq 1$\\ $R_1 = R_2 = R_3$\end{tabular}                                                         & \begin{tabular}[c]{@{}c@{}}$(1|2,3)$\\ $(2|1,3)$\\ $(3|-)$\\$ (e|3)$\end{tabular}                   & \begin{tabular}[c]{@{}c@{}}$R_1 = R_2$ \\$ R_2 + R_3 \leq 1$\end{tabular}                               \\ \hline
\begin{tabular}[c]{@{}c@{}}$(1|3)$\\$ (2|3)$\\$ (3|1,2)$\\ $(e|-)$\end{tabular}                     & \begin{tabular}[c]{@{}c@{}}$R_1 + R_2 \leq 1$\\$ R_1 \leq R_3$\\$ R_2 \leq R_3$\\$ R_3 \leq R_1 + R_2$\end{tabular}                               & \begin{tabular}[c]{@{}c@{}}$(1|3)$\\ $(2|3)$\\$ (3|1,2)$\\$ (e|1)$\end{tabular}                     & \begin{tabular}[c]{@{}c@{}}$R_2 = R_3$\\$ R_1 + R_3 \leq 1$\end{tabular}                                \\ \hline
\begin{tabular}[c]{@{}c@{}}$(1|3)$\\ $(2|3)$\\$ (3|1,2)$\\$ (e|2)$\end{tabular}                     & \begin{tabular}[c]{@{}c@{}}$R_1 = R_3$\\ $R_2 + R_3 \leq 1$\end{tabular}                                                              & \begin{tabular}[c]{@{}c@{}}$(1|3)$\\ $(2|1)$\\$ (3|1,2)$\\ $(e|-)$\end{tabular}                     & \begin{tabular}[c]{@{}c@{}}$R_1 = R_3$\\ $R_2 + R_3 \leq 1$\\ $R_2 \leq R_1$\end{tabular}                    \\ \hline
\begin{tabular}[c]{@{}c@{}}$(1|3)$\\ $(2|1)$\\ $(3|1,2)$\\ $(e|2)$\end{tabular}                     & \begin{tabular}[c]{@{}c@{}}$R_1 = R_3$\\$ R_2 + R_3 \leq 1$\end{tabular}                                                              & \begin{tabular}[c]{@{}c@{}}$(1|3)$\\ $(2|1,3)$\\ $(3|1)$\\ $(e|-)$\end{tabular}                     & \begin{tabular}[c]{@{}c@{}}$R_1 + R_2 \leq 1$\\ $R_1 = R_3$\\$ R_2 \leq R_3$\end{tabular}                    \\ \hline
\begin{tabular}[c]{@{}c@{}}$(1|3)$\\ $(2|1,3)$\\ $(3|1)$\\ $(e|2)$\end{tabular}                     & \begin{tabular}[c]{@{}c@{}}$R_1 + R_2 \leq 1$\\ $R_1 = R_3$\end{tabular}                                                              & \begin{tabular}[c]{@{}c@{}}$(1|2,3)$\\$ (2|1,3)$\\ $(3|1)$\\ $(e|-)$\end{tabular}                   & \begin{tabular}[c]{@{}c@{}}$R_3 \leq R_1$\\$ R_2 \leq R_1$\\ $R_2 + R_3 \leq 1$\\$ R_1 \leq R_2 + R_3$\end{tabular} \\ \hline
\begin{tabular}[c]{@{}c@{}}$(1|2,3)$\\ $(2|1,3)$\\$ (3|1)$\\$ (e|2)$\end{tabular}                   & \begin{tabular}[c]{@{}c@{}}$R_1 = R_3$\\$ R_2 + R_3 \leq 1$\end{tabular}                                                              & \begin{tabular}[c]{@{}c@{}}$(1|2,3)$\\$ (2|1,3)$\\$ (3|1)$\\ $(e|3)$\end{tabular}                   & \begin{tabular}[c]{@{}c@{}}$R_1 = R_2$\\$ R_2 + R_3 \leq 1$\end{tabular}                                \\ \hline
\begin{tabular}[c]{@{}c@{}}$(1|2,3)$\\$ (2|1,3)$\\ $(3|1,2)$\\$ (e|-)$\end{tabular}                 & \begin{tabular}[c]{@{}c@{}}$R_1 \leq 1$\\$ R_2 \leq 1$\\$ R_3 \leq 1$\\$ R_1 \leq R_2 + R_3$\\$ R_2 \leq R_1 + R_3$\\$ R_3 \leq R_1 + R_2$\end{tabular} & \begin{tabular}[c]{@{}c@{}}$(1|2,3)$\\$ (2|1,3)$\\$ (3|1,2)$\\$ (e|1)$\end{tabular}                 & \begin{tabular}[c]{@{}c@{}}$R_1 \leq 1$\\ $R_2 \leq 1$\\ $R_2 = R_3$\end{tabular}                          \\ \hline
\begin{tabular}[c]{@{}c@{}}$(1|2,3)$\\ $(2|1,3)$\\$ (3|1,2)$\\ $(e|2)$\end{tabular}                 & \begin{tabular}[c]{@{}c@{}}$R_2 \leq 1$\\$ R_3 \leq 1$\\$ R_1 = R_3$\end{tabular}                                                        & \begin{tabular}[c]{@{}c@{}}$(1|2,3)$\\$ (2|1,3)$\\ $(3|1,2)$\\$ (e|3)$\end{tabular}                 & \begin{tabular}[c]{@{}c@{}}$R_1 \leq 1$\\ $R_3 \leq 1$\\ $R_1 = R_2$\end{tabular}                          \\ \hline
\end{tabular}

\end{table}

\newcommand{\norm}[1]{\left\lVert#1\right\rVert}
\section{Proof of Proposition \ref{theo:SCOMPenhanced}} \label{proof:SCOMP}
Observe that the first three rate constraints of Proposition \ref{theo:SCOMPenhanced} follow from the achievability proof of~\cite{liu2017capacity}. We are done if we show that \eqref{eq:sec_comp} implies for each $\Dv\in\Delta$,
\begin{align}
I(\Xv_i(\Dv); (W_{K,\Dv}: K\in T)| \Xv_{A_e}(\Dv)) = o(1) \label{eqn-DvIbnd}
\end{align}
as $r \rightarrow \infty$. Since different parts of messages are independently encoded, the above would also ensure that 
\begin{align*}
I(\Xv_i; (W_{K,\Dv}: K\in T, \Dv\in\Delta )| \Xv_{A_e}) = o(1) \textrm{ as } r\rightarrow \infty.
\end{align*}
To show that \eqref{eqn-DvIbnd} holds under \eqref{eq:sec_comp}, we pick $\Dv\in \Delta$ and focus on only the part of message conveyed via this choice of decoding configuration. For the remainder of proof, we drop the reference to $\Dv$ remembering that $\Xv_j$, $R_j$, $S_K$ stand for $\Xv_j(\Dv)$, $R_j(\Dv)$, $S_K(\Dv)$, respectively.

Recall that $\Xv_j \in [2^{rR_j}]$, $j \in [n]$, are independent and uniformly distributed. Let $Z_i= \{i\}\cup A_e$ for some $i \in A_e^c$. Set $T = \{K: K \subseteq [n], K \not\subseteq A_e\}$ and $P_Q= \bigcup_{J \in Q} J \backslash A_e$, for any $Q \subseteq T$.  

Define $R_{Z_i} = R_i + \sum_{j \in A_e} R_j$ and $R_K = \sum_{j \in K} R_j$, $K \in T$. We have $\Xv_{Z_i} = (\Xv_i, \Xv_{A_e}) \in [2^{rR_i}]  \times \prod_{j\in A_e} [2^{rR_j}] = [2^{rR_{Z_i}}]$ and for each $K \in T$, $\Xv_K = (x_i, i\in K) \in \prod_{j\in K} [2^{rR_j}] = [2^{rR_K}]$.  Then $((\Xv_{K}, K \in T), \Xv_{Z_i})$ is a well-defined discrete memoryless correlated source.

For each  $K \in T$, the random mapping $b_K: [2^{rR_K}] \rightarrow [2^{rS_K}]$ uniformly and independently maps each sequence $\xv_K$ to  $w_K \in [2^{rS_K}]$. Denote by $B_K$ and $W_K$ the random encoding and the random bin index corresponding to $b_K$ and $w_K$, respectively. Let $w_T = (w_K, K \in T)$, $W_T = (W_K, K \in T)$, $b_T = (b_K, K \in T)$ and $B_T = (B_K, K \in T)$. Note that for each $Q \subseteq T$ 
\begin{align}
\sum_{J \in Q} S_J = \sum_{\substack{K \subseteq P_Q\cup A_e\\ K \not\subseteq A_e}} S_K,
\end{align}
and for each $Q \subseteq T$ and $Z_i$,
\begin{align*}
H( \Xv_{\bigcup_{J\in Q}}|\Xv_{Z_i}) &= H(\Xv_{\bigcup_{J\in Q}}|\Xv_i \cup \Xv_{A_e})= \sum_{j \in (P_Q\backslash \{i\})} R_j.
\end{align*}

Using our notation, Theorem 1 in \cite{6283010} can be restated as:

\begin{theorem} \label{theo:RB}
If for each $Q \subseteq T$, 
\begin{align}
\sum_{\substack{K \subseteq P_Q\cup A_e\\K \not\subseteq A_e}} S_K <  \sum_{j \in (P_Q\backslash \{i\})} R_j.
\end{align}
then
\begin{align}
\label{eq:th1_prob} \mathbb{E}_{B_T}\norm{p_{\Xv_{Z_i}W_T} - p_{\Xv_{Z_i}}\prod_{K\in T} p^U_{[2^{rS_K}]}}_{1} < o(2^{-\beta r}),
\end{align}
\end{theorem}
where: (1) The outer expectation is only over the choice of random binning; (2) $p_{\Xv_{Z_i}W_T}$ is the joint pmf induced by a particular random binning; (3) $p^U_{\mathcal A}$ is the uniform distribution over set $\mathcal A$,  and (4) $\beta >0$ is the rate of convergence. 
Since \eqref{eq:th1_prob} exponentially converges to 0 as $r$ goes to infinity  \cite{6283010}, there exists a sequence of binning schemes $\{b^*_{T,r}: r\in\mathbb{N}\}$ such that the sequence of joint pmfs $\{p^{*(r)}_{\Xv_{Z_i}W_T}: r\in\mathbb{N}\}$ induced by the sequence of  binning schemes satisfies
\begin{align}
\label{eq:th1_eq2}  \norm{p^{*(r)}_{\Xv_{Z_i}W_T} - p_{\Xv_{Z_i}}\prod_{K\in T} p^U_{[2^{rS_K}]}}_{1} < o(2^{-\beta r}),
\end{align}

Now, we use Lemma 2.7 from \cite{csiszar2011information}:
\begin{lemma} \label{theo:lemma2_7}
If $p_1$ and $p_2$ are two distributions over a finite set $\Xv$ such that $\sum\limits_{\xv \in \Xv} \norm{p_1(\xv) - p_2(\xv)}_{1} \leq \theta \leq \frac{1}{2}$, then
\begin{align}
\norm{H(p_1) - H(p_2)}_1 \leq -\theta \log{\frac{\theta}{|\Xv|}}.
\end{align}
\end{lemma}

Recall $ |\Xv| = r(\sum_{j \in \{i\} \cup A_e}R_j + \sum_{K \in T} S_K) = \alpha r$.  Thus, by invoking Lemma \ref{theo:lemma2_7} for the two distributions in \eqref{eq:th1_eq2}, we see that for the sequence of binning schemes $\{b^*_{T,r}: r\in\mathbb{N}\}$:
\begin{align*}
I(\Xv_i \cup \Xv_{A_e} ; (W_K, K \in T))& < o(\alpha r2^{-\beta r})\\
\Rightarrow\quad I(\Xv_i ;(W_K, K \in T) | \Xv_{A_e}) &< o(\alpha r2^{-\beta r}).
\end{align*}
Therefore, the security condition \eqref{final_secure_eq2} holds as $r \rightarrow \infty$.

\ifitw
\else
\section{Proof of Theorem \ref{Prop_hg_equi}} \label{proof:g_to_h}
To prove $\mathcal{R}_{\mathrm{g}} \subseteq  \mathcal{R}_{\mathrm{h}}$, suppose $(R_1, \dots, R_n) \in  \mathcal{R}_{\mathrm{g}}$. Then there exists $g(J), J \subseteq [n]$, satisfying \eqref{pm_a_1}-\eqref{pm_a_5} such that
\begin{align}
R_i \leq g(B_i \cup \{i\}) - g(B_i).
\end{align}
Define a set function $h: 2^{\{0\} \cup [n]} \rightarrow R_{\geq 0}$ as in \eqref{g_to_h_1}-\eqref{g_to_h_4}. To prove \eqref{SH_rate}, we use \eqref{g_to_h_2} and \eqref{g_to_h_4} as 
\begin{align}
h(\{0\}) &= h(\emptyset)+\frac{g([n])}{\sum_{i=1}^n g(\{i\})}\\
 &= \frac{1}{\sum_{i=1}^n g(\{i\})} = 
\frac{h(\{i\})}{g(\{i\})}.
\end{align}

Hence, we have
\begin{align}
R_i &\leq g(B_i \cup\{i\}) - g(B_i) = g(\{i\}) = \frac{h(\{i\})}{h(\{0\})}.
\end{align}

Constraints \eqref{SH_1} and \eqref{SH_2} follow directly from the definition. To derive \eqref{SH_4}, we have
\begin{align}
h(N) &= h([n]\cup \{0\}) = h([n]) +\frac{ g(\emptyset)}{\sum_{i=1}^n g(\{i\})}\\
&  = h([n])= \frac{\sum_{i=1}^n g(\{i\})}{\sum_{i=1}^n g(\{i\})} =1,
\end{align}
from which $h(N \backslash \{i\}) = h([n]) =1$ follows trivially, and for $i \in [n]$ we have 
\begin{align}
 h(N \backslash \{i\}) &= h(([n] \backslash \{i\}) \cup \{0\})\\
& = \frac{\sum_{j \neq i} g(\{j\}) + g(\{i\})}{\sum_{k=1}^n g(k)} = 1.
\end{align}

Constraint \eqref{SH_5} can be derived from \eqref{pm_a_5} and  \eqref{g_to_h_4} as follows:
\begin{align}
h(\{i\} \cup A_i \cup \{0\}) &= h(\{i\} \cup A_i) +\frac{ g(B_i)}{\sum_{i=1}^n g(\{i\})}  \\
& =h(\{i\} \cup A_i) + \frac{ g(B_i \cup \{i\}) - g(\{i\})}{\sum_{i=1}^n g(\{i\})}  \\
& =h(A_i) + \frac{ g(B_i \cup \{i\})}{\sum_{i=1}^n g(\{i\})}  \\
&= h(A_i \cup \{0\}).
\end{align}

To prove \eqref{SH_6}, we consider the following cases.
\begin{itemize}
\item When $ 0 \not\in J$ and $i \neq 0$, from the definition of $h$ and non-negativity of $g$, we have
\begin{align}
h(J) = \sum_{j\in J} h(j) \leq \sum_{j \in J \cup \{i\}} h(j).
\end{align}
\item When $ 0 \in J$ and $i \neq 0$, let $J = K \cup \{0\}, 0 \not\in K $. Then,
\begin{align}
h(J) &= h(K \cup \{0\}) \\
&= h(K) +\frac{g(\overline{K})}{\sum_{i=1}^n g(\{i\})},
\end{align}
and
\begin{align}
h(J \cup \{i\}) &= h(K \cup \{i\} \cup \{0\}) \\
&=h(K \cup \{i\}) + \frac{ g(\overline{K} \backslash \{i\})}{\sum_{i=1}^n g(\{i\})}.
\end{align}
Subtracting one from another gives
\begin{align}
&h(J) - h(J \cup \{i\})\\
 &= h(K) - h(K \cup \{i\}) +\frac{g(\overline{K}) - g(\overline{K} \backslash \{i\})}{\sum_{i=1}^n g(\{i\})} \\
&= \frac{g(\overline{K}) - g(\{i\}) - g(\overline{K} \backslash \{i\})}{\sum_{i=1}^n g(\{i\})}\\
\label{h_J_i_proof} &\leq 0,
\end{align}
where \eqref{h_J_i_proof} follows from the submodularity of $g$ in \eqref{pm_a_4}.
\item When  $ 0 \not\in J$ and $i = 0$,
\begin{align}
h(J) &= \frac{\sum_{j \in J} g(\{j\})}{\sum_{j\in [n]}g(\{j\})}, \\
h(J \cup \{i\}) & = \frac{\sum_{j\in J} g(\{j\}) +g(\overline{J})}{\sum_{j\in [n]} g(\{j\})}\geq h(J).
\end{align}
\end{itemize}
Hence, \eqref{SH_6} holds. Finally, \eqref{SH_7} is proved by considering the following cases. First, for compact notation, let $$h(J) + h(J \cup \{i,k\}) = \frac{h'}{\sum_{j\in [n]} g(\{j\})},$$ and  $$h(J \cup \{i\}) + h(J \cup \{k\}) = \frac{h''}{\sum_{j\in [n]} g(\{j\})}.$$
\begin{itemize}
\item When $0 \not\in J \text{ and } i,k \neq 0$, \eqref{SH_7} is satisfied due to submodularity of function $g$.
\item When $0 \in J \text{ and } i,k \neq 0$, define $J = K \cup \{0\}$ and $ L = \overline{K} \backslash \{i,k\}$.  Then,
\begin{align*}
h' &= \sum_{j\in K} g(\{j\}) +g(\overline{K}) + \sum_{l\in K} g(\{l\}) \notag\\
&~~~~~~~~~~~~~~~~~~~~+ g(\{i\}) + g(\{k\}) + g(L),
\end{align*}
and 
\begin{align*}
h'' = & \sum_{j \in K} g(\{j\})  + g(\{i\}) + g(\overline{K} \backslash \{i\}) \notag \\
&~~~~~~~~~~~~~~~+ \sum_{l \in K} g(\{l\}) + g(\{k\}) + g(\overline{K} \backslash \{k\}).
\end{align*}
Due to submodularity of $g$, we have
\begin{align*}
&g(\overline{K}) + g(\overline{K}\backslash \{i,k\}) - g(\overline{K} \backslash \{i\}) - g(\overline{K} \backslash \{k\})  =\\
&g(L \cup \{i,k\}) + g(L) - g(L\cup \{i\}) - g(L \cup \{k\})  \leq 0.
\end{align*}
Hence,
\begin{align*}
h(J) + h(J \cup \{i,k\})-h(J \cup \{i\}) - h(J \cup \{k\}) \leq 0 .
\end{align*}
\item When $ 0 \not\in J, i = 0, k \neq 0$, we have
\begin{align*}
h'&= \sum_{j\in J} g(\{j\}) + \sum_{j\in J} g(\{j\}) + g(\{k\}) + g(\overline{J} \backslash \{k\}), \\
h'' &= \sum_{j\in J} g(\{j\}) + g(\overline{J}) + \sum_{j\in J} g(\{j\}) + g(\{k\}).
\end{align*}
Since $g$ is monotonic, we have
\begin{align*}
g(\overline{J} \backslash \{k\}) - g(\{k\}) \leq 0.
\end{align*}
Hence,
\begin{align*}
h(J) + h(J \cup \{i,k\})-h(J \cup \{i\}) - h(J \cup \{k\}) \leq 0 .
\end{align*}
The  case $ 0 \not\in J, k = 0, i \neq 0$ follows from symmetry.
\end{itemize}

Next, we prove $\mathcal{R}_{\mathrm{h}} \subseteq  \mathcal{R}_{\mathrm{g}}$. 

Suppose $(R_1, \dots, R_n) \in  \mathcal{R}_{\mathrm{h}}$, then there exists $h(J), J \subseteq \{0\} \cup [n]$, satisfying \eqref{SH_1}-\eqref{SH_7} such that
\begin{align}
R_i \leq \frac{h(\{i\})}{h(\{0\})}, ~~~ i\in [n].
\end{align}
Define a set function $g: 2^{[n]} \rightarrow [0,1]$ as in \eqref{h_to_g}. We prove \eqref{pm_a_0} using \eqref{SH_5} and the definition of $g$:
\begin{align}
R_i &\leq \frac{h(\{i\})}{h(\{0\})} \\
&= \frac{1}{h(\{0\})} (-h(A_i) + h(\{i\} \cup A_i)) \\
&= \frac{1}{h(\{0\})} (h(A_i \cup \{0\}) -h(A_i) -  \notag\\
&~~~~~~~~~~~~~~~ (h(A_i \cup \{0\}) - h(\{i\} \cup A_i))) \\
&= \frac{1}{h(\{0\})} (h(A_i \cup \{0\}) - h(A_i) -  \notag\\
&~~~~~~~~~~~~~~~ (h(\{i\} \cup A_i \cup \{0\}) - h(\{i\} \cup A_i))) \\
&= g(\{i\} \cup B_i) - g(B_i).
\end{align}

Next, we prove for this choice of $g$, that constraints \eqref{pm_a_1}-\eqref{pm_a_5} are satisfied.

Due to \eqref{SH_4}, constraint \eqref{pm_a_1} follows immediately as
\begin{align}
g(\emptyset) &= \frac{h([n] \cup \{0\}) -h([n])}{h(\{0\})} = 0.
\end{align}

Constraint \eqref{pm_a_2} also follows from definition
\begin{align}
g([n]) &= \frac{h(\{0\}) - h(\emptyset)}{h(\{0\})}= 1.
\end{align}

To prove the monotonicity of $g$, i.e.,  \eqref{pm_a_3}, we first note that
\begin{align}
g(J) &= \frac{h(\overline{J} \cup \{0\}) - h(\overline{J})}{h(\{0\})}, \\
g(J \cup \{i\}) &= \frac{h(\overline{J}\backslash \{i\} \cup \{0\}) - h(\overline{J} \backslash \{i\})}{h(\{0\})}.
\end{align}
Denote $\overline{J} \backslash \{i\} = K$. By  taking the difference between $g(J)$  and $g(J \cup \{i\})$, we see that
\begin{align*}
& g(J) - g(J \cup \{i\})\\
&= \frac{h(K) + h(K \cup \{i,0\}) - h(K \cup \{i\}) - h(K \cup \{0\})}{h(\{0\})}\leq 0,
\end{align*}
where the above follows from submodularity of $h$ in \eqref{SH_7}.

To prove the submodularity of $g$, i.e., \eqref{pm_a_4},  define $L = \overline{J} \backslash \{i,k\}$. Then, we have
\begin{align*}
g(J)  &= \frac{h(L\cup \{k\} \cup \{i\} \cup \{0\}) - h(L \cup \{k\} \cup \{i\})}{h(\{0\})} \\
& = \frac{h(L\cup \{k\} \cup \{i\} \cup \{0\}) - h(L) - h(\{i\}) - h(\{k\})}{h(\{0\})},
\end{align*}
and
\begin{align*}
g(J \cup \{i\} \cup \{k\}) &=  \frac{h(L\cup \{0\}) - h(L)}{h(\{0\})},\\
g(J \cup \{i\} ) &=  \frac{h(L \cup \{k\} \cup \{0\}) - h(L \cup \{k\} )}{h(\{0\})}\\
&= \frac{h(L\cup \{k\} \cup \{0\}) - h(L) - h(\{k\})}{h(\{0\})}, \\
g(J \cup \{k\} ) &=  \frac{h(L \cup \{i\} \cup \{0\}) - h(L \cup \{i\} )}{h(\{0\})} \\
&= \frac{h(L \cup \{k\} \cup \{0\}) - h(L) - h(\{i\})}{h(\{0\})}.
\end{align*}
Taking the difference and using the submodularity of $h$, we see that
\begin{align*}
&g(J) + g(J \cup \{i\} \cup \{k\}) -  g(J \cup \{i\} )  -g(J \cup \{k\})\\
& = \frac{1}{ (h(\{0\})}\big(h(L \cup \{k\} \cup \{i\} \cup \{0\}) + h(L \cup \{0\}) \notag \\
&~~~~~~~~~~~~~ - h(L \cup \{k\} \cup \{0\}) -h(L \cup \{i\} \cup \{0\})) 
\leq 0.
\end{align*}

Finally, to prove \eqref{pm_a_5}, we write
\begin{align}
&g(\{i\} \cup B_i) - g(B_i) \\
\nonumber&= \frac{h(A_i \cup \{0\}) - h(A_i) - h(\{i\} \cup A_i \cup \{0\}) + h(\{i\} \cup A_i)}{h(\{0\}} \\
\label{pma5_1} & = \frac{h(\{i\} \cup A_i)- h(A_i)}{h(\{0\})} \\
&= \frac{h(\{i\})}{h(\{0\})} \\
&= \frac{h([n])- h([n] \backslash \{i\})}{h(\{0\})} \\
\label{pma5_2} &= \frac{h([n]\backslash \{i\} \cup \{0\})- h([n] \backslash \{i\})}{h(\{0\})} = g(\{i\}).
\end{align}
Note that \eqref{pma5_1} is due to \eqref{SH_5} and \eqref{pma5_2} is due to \eqref{SH_4}.

Therefore, $\mathcal{R}_{\mathrm{g}} = \mathcal{R}_{\mathrm{h}}$.
\fi

\end{document}